# Development and Evaluation of an Ontology for Non-Invasive Respiratory Support in Acute Care

*Md Fantacher Islam[1¶], Jarrod Mosier[2¶], Vignesh Subbian[1¶]

[1]College of Engineering, The University of Arizona, Tucson, Arizona, United States

[2]College of Medicine, The University of Arizona, Tucson, Arizona, United States

*Corresponding Author

E-mail: fantacher@arizona.edu

[¶]These authors contributed equally to this work.

## Abstract

Managing patients with respiratory failure increasingly involves non-invasive respiratory support (NIRS) strategies to support respiration, often preventing the need for invasive mechanical ventilation. However, despite the rapidly expanding use of NIRS, there remains a significant challenge to its optimal use across all medical circumstances. It lacks a unified ontological structure, complicating guidance on NIRS modalities across healthcare systems. In this study, wes introduced NIRS ontology to support knowledge representation in acute care settings by providing a unified framework that enhances data clarity and interoperability, laying the groundwork to support future clinical decision-making. We developed NIRS ontology using the Web Ontology Language (OWL) and Protégé to organize clinical concepts and relationships. To enable rule-based clinical reasoning beyond hierarchical structures, we added Semantic Web Rule Language (SWRL) rules. We evaluated logical reasoning by adding a sample of 6 patient scenarios. We used SPARQL queries to retrieve and test targeted inferences. The ontology has 145 classes, 11 object properties, and 18 data properties across 949 axioms that establish concept relationships. To standardize clinical concepts, we added 392 annotations, including descriptive definitions based on controlled vocabularies. SPARQL queries successfully validated all test cases (rules) by retrieving appropriate patients' outcomes: for instance, a patient treated with HFNC (high-flow nasal cannula) for 2 hours due to acute respiratory failure may avoid endotracheal intubation. This NIRS ontology formally represents domain-specific concepts, including ventilation modalities, patient characteristics, therapy parameters, and outcomes. SPARQL query evaluations across clinical scenarios confirmed the ontology's ability to support rule-based reasoning and therapy recommendations, providing a foundation for consistent documentation practices, integration into clinical data models, and advanced analysis of NIRS outcomes. In conclusion, we unified NIRS concepts into an ontological framework and demonstrated its applicability through the evaluation of patient scenarios and alignment with standardized vocabularies.

# Background

Noninvasive respiratory support (NIRS) modalities have emerged as effective strategies for managing patients with acute respiratory failure, such as acute exacerbations of chronic obstructive pulmonary disease), cardiogenic pulmonary edema, and acute hypoxemic respiratory failure (AHRF) (1–3). These strategies have expanded available respiratory support options beyond conventional oxygen or invasive mechanical ventilation (4). Despite their clinical effectiveness, NIRS modalities such as Continuous Positive Airway Pressure (CPAP) and High-Flow Nasal Cannula (HFNC) originated in different contexts (CPAP for sleep medicine (5) and HFNC for neonatal care (6)) before being adopted in adult acute care settings (7). Historically, these modalities were studied and documented separately, but the COVID-19 pandemic unified them into NIRS interventions for the massive influx of hypoxemic patients (8). Consequently, data captured within Electronic Health Records (EHRs) tends to be fragmented, with each modality documented using distinct parameter sets, creating interoperability issues (9). Nonetheless, the lack of standardized representations regarding the nature and use of NIRS too often limits reproducible cohort definitions, phenotyping, and cross-site comparability across the spectrum of heterogeneous respiratory failure conditions (10).

More specifically, a key challenge for clinician-scientists, research informaticians, and administrators is the lack of standardized methods for describing and categorizing NIRS modalities and their combinations, which complicates collecting and analyzing large-scale NIRS data (11). This highlights challenges in using secondary EHR data, which are often heterogeneous and lack explicit domain knowledge, making interpretation difficult due to inconsistency, inaccuracy, and incompleteness (12). These varying documentation practices within and between healthcare systems limit the ability to further real-world use and evidence related to the effectiveness and safety of NIRS (13). These multifaceted challenges highlight the need for a domain-specific ontology and knowledge representation framework for NIRS to facilitate accurate information exchange and inform clinical decision-making across healthcare settings (14). Recent studies have developed taxonomies to clarify complex terminologies, establish consistent vocabularies, and support more informed comparisons related to invasive mechanical ventilation(15,16). Similarly, while domain-specific ontologies have been developed to standardize intensive care diagnoses, these frameworks do not capture the granular therapeutic parameters required for NIRS phenotyping (17). Yet, a unified standardization approach specific to NIRS is missing.

While standard taxonomies provide essential lists of terms, they lack the semantic reasoning required to handle imperfect or conflicting data. Addressing these gaps through ontology-based approaches offers a promising solution for enhancing interoperability, enabling more advanced clinical queries, and providing a structured foundation for improving NIRS practices (18). Ontologies provide formal, machine-interpretable representations that unify clinical data by explicitly defining domain-specific concepts and relationships, supporting semantic clarity and enabling meaningful data integration (14). Although prior ontology efforts have addressed mechanical ventilation and temporal relationships in clinical events, such as the Time Event Ontology (TEO), which uses Allen's interval algebra, a comprehensive ontology tailored to NIRS has yet to be developed (19).

The goal of this study is to develop a domain-specific ontology for Non-Invasive Respiratory Support (NIRS). Specifically, we aim to: (1) formalize the hierarchical relationships between distinct NIRS modalities, devices, and their parameters, (2) map these concepts to standard controlled vocabularies (e.g., SNOMED-CT, ICD-10, and MedDRA) to ensure interoperability, and (3) demonstrate the ontology's applicability in clinical phenotyping.

# Methods

**Fig 1** illustrates the end-to-end workflow, beginning with incorporating domain knowledge and extracting data elements from the eICU (Electronic Intensive Care Unit) collaborative research database, continuing through modeling in Protégé, and concluding with creating SWRL (Semantic Web Rule Language) rules and SPARQL queries to support competency-based insights (20).

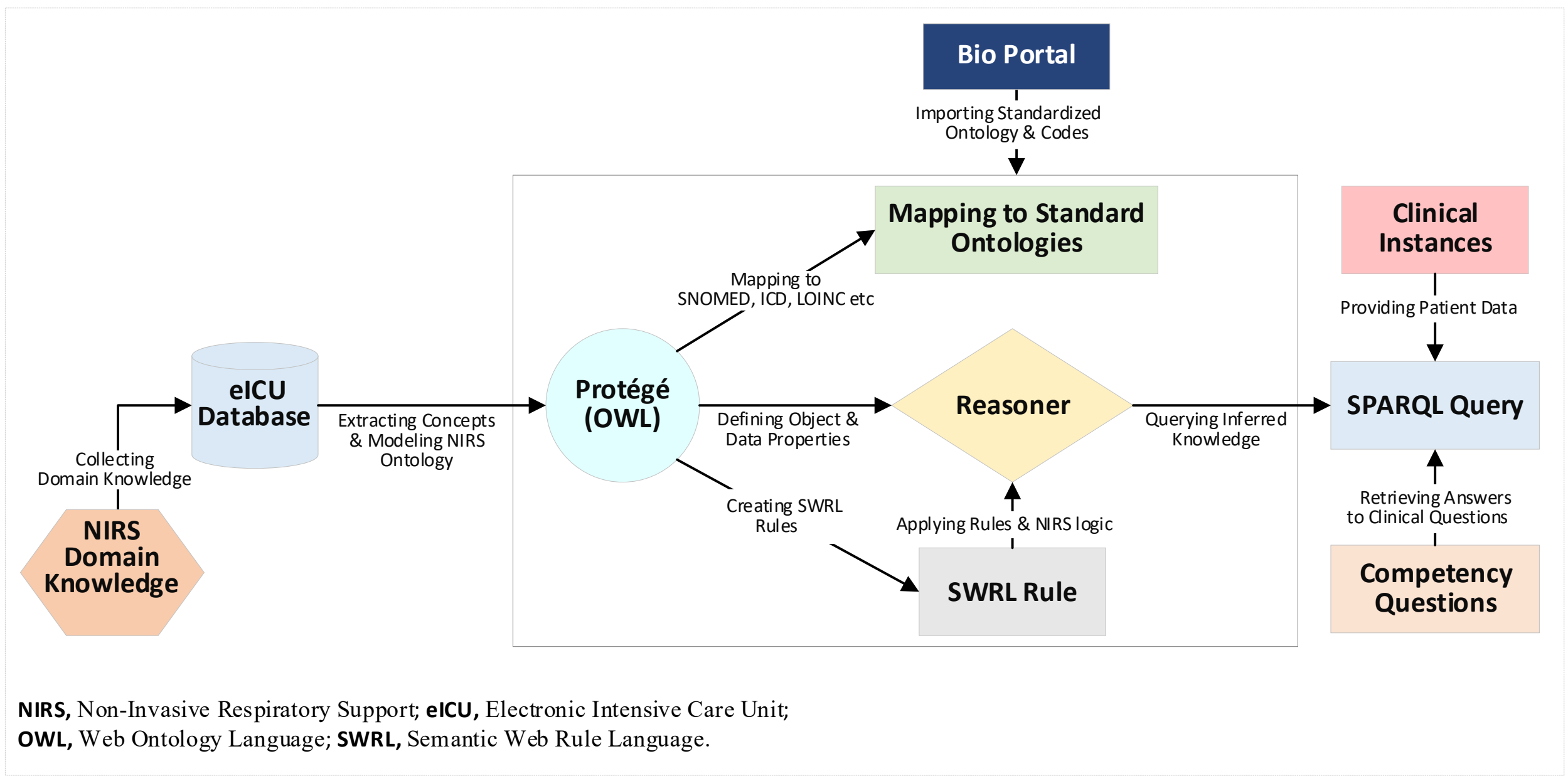

**Fig 1** Workflow of the NIRS Ontology Development and Application.

## Data Source

A part of this study involved the analysis of a publicly available eICU database v2.0, containing de-identified electronic health record data from over 200,000 admissions across 335 units at 208

United States Hospitals between 2014 and 2015 (20). This study is exempt from institutional review board (IRB) approval due to the retrospective design, lack of direct patient intervention, and the security schema, for which the re-identification risk was certified as meeting safe harbor standards by an independent privacy expert (Privacert, Cambridge, MA) (Health Insurance Portability and Accountability Act Certification no. 1031219-2).

## Data Extraction and Clinical Concepts for NIRS

We extracted clinical concepts and data on conditions commonly treated with NIRS modalities from five core tables in the eICU database: *respiratoryCharting*, *carePlanGeneral*, *diagnosis*, *nurseCharting*, and *treatment*. These tables provided key factors, including NIRS indications, modality, and settings. A review of published research informed our inclusion criteria, which aimed to cover all acute illnesses (e.g., sepsis, septic shock, pneumonia) (21,22) and decompensated chronic illnesses (e.g., cardiogenic pulmonary edema, COPD (chronic obstructive pulmonary disease), and Obesity Hypoventilation Syndrome) typically managed with NIRS (23,24). We also extracted data known to be either comorbidities or confounders associated with improved or worsened outcomes with NIRS (25–27). Specific threshold values for data extraction (e.g., BMI cutoffs) and reasoning rules were selected based on clinical practice guidelines (such as the American Thoracic Society (ATS)) and peer-reviewed articles to ensure clinical relevance. To ensure accuracy, an intensivist reviewed and validated the underlying domain knowledge, including class hierarchies and safety thresholds to ensure alignment with real-world acute care workflows.

## Ontology Construction and Mapping

We constructed the NIRS ontology using Web Ontology Language (OWL) semantics and Protégé software to organize clinical concepts and relationships (**Fig 1**). We defined core superclasses and subclasses for NIRS strategies by reviewing published literature, collaborating with an intensivist through a formal iterative process, and analyzing the eICU database. This approach ensured that the ontology is intuitive for human users and interoperable with computational tools, including Protégé. To enhance semantic interoperability, we added annotations to leaf or terminal classes. For defining the concepts, we use 'rdfs:label' for readable names (e.g., "Chronic Obstructive Pulmonary Disease" for COPD), 'rdfs:comment' for a concise clinical definition, and 'rdfs:seeAlso' to link to resources such as BioPortal. To map concepts to standard vocabularies, we created a custom annotation 'hasOntoCode,' which holds vocabulary codes with IDs, such as SNOMED-CT: 13645005. A team of two informaticians performed these mappings through a two-round iterative process. The initial mapping was conducted using the BioPortal tool. We defined object and data properties to interrelate classes or subclasses and store associated values. Object properties establish relationships between classes (e.g., linking a patient's indication to a therapy), while data properties capture parameters such as oxygen flow rate or $FiO_2$, with domains and ranges set for validation.

## Reasoning and Inference in the NIRS Ontology

After developing the NIRS hierarchy and annotations, we tested the logical reasoning using hypothetical patient records (instances or individuals) that represented real-world clinical scenarios. For instance, we would assign a patient with acute respiratory distress as HFNC (high-flow nasal cannula), with parameters such as flow rate and $FiO_2$, or noninvasive positive pressure ventilation (NIPPV), with IPAP (inspiratory positive airway pressure) and EPAP (expiratory positive airway pressure). We used the Pellet reasoner for checking consistency and inferring relationships and class memberships using OWL axioms. SWRL rules added domain-specific logic, enhancing clinical reasoning beyond basic hierarchies. These rules, created in Protégé, enabled complex decision-making, such as identifying critical conditions or suggesting therapies. SPARQL queries retrieved targeted data, allowing the analysis of trends such as therapy selection or outcomes. Together, these components formed a robust ontological framework for representing clinically meaningful knowledge, enabling semantic reasoning tasks such as classification and inference, and supporting integration of heterogeneous data sources for clinical decision-making and analysis.

# Ontology Framework for NIRS

The NIRS ontology provides a structured representation of clinical knowledge related to non-invasive respiratory support, capturing the conditions that indicate its use, the specific modalities and parameters of therapy administered, and the observed patient responses or outcomes.

## Structural Design of the Ontology

The NIRS ontology's hierarchy begins with 'Thing' as the top-level class (**Fig 2**), serving as the universal root from which we derive primary superclasses: *Therapy*, *Indication*, *Patient*, and *Outcome*. The 'Therapy' superclass includes the required therapy details classes, such as the 'TherapeuticPhase' subclass, which captures the clinical intent of therapies, as defined by the ATS clinical practice guidelines (28–30). This superclass also has other subclasses of device types, parameters, and timing, encompassing interventions, such as HFNC systems, positive airway pressure variants, and therapy duration. The ‘Patient’ superclass includes demographics, comorbidities, and other key patient characteristics (see Interactive Class Hierarchy).

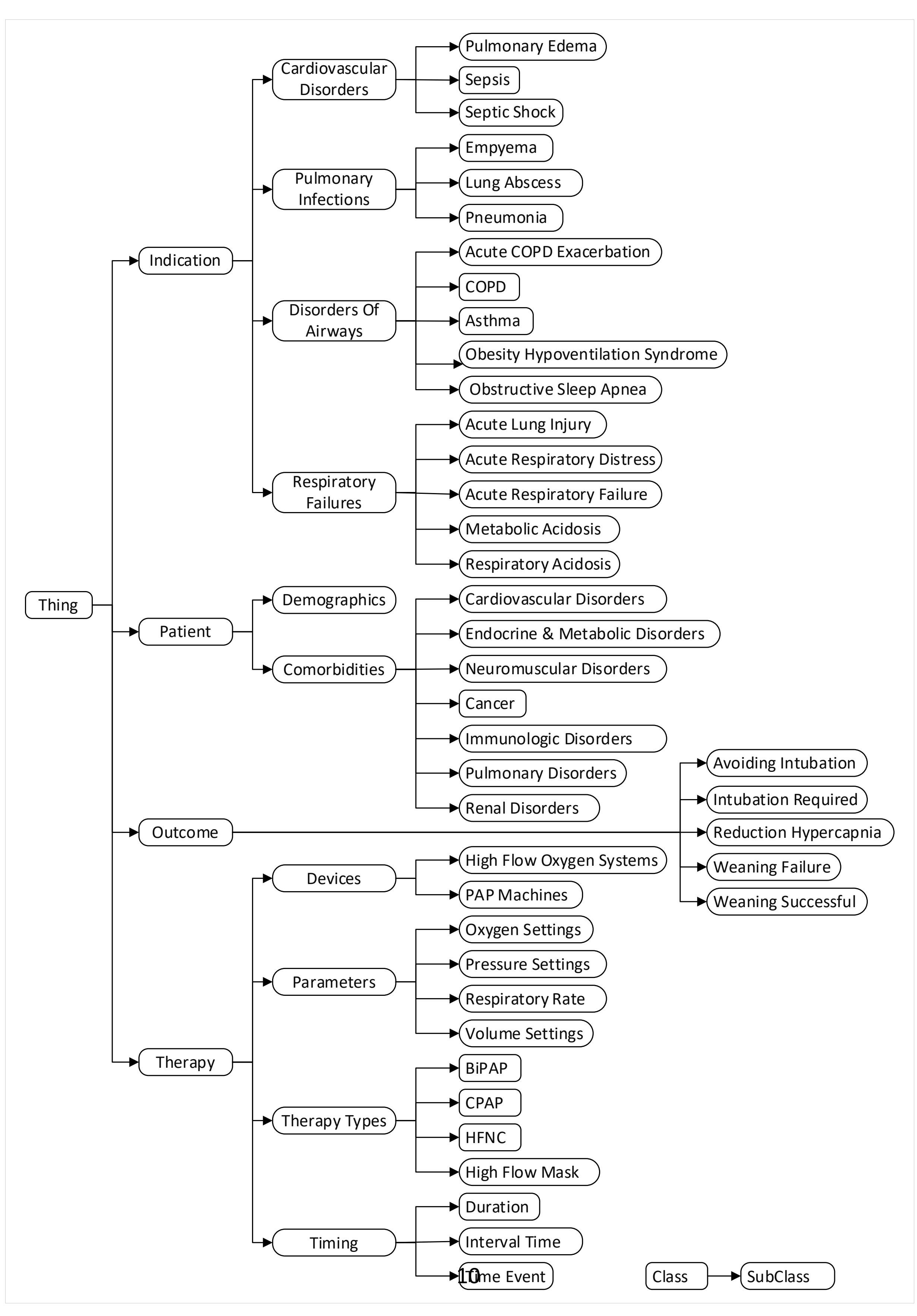

Thing
Indication
Cardiovascular Disorders
Pulmonary Edema
Sepsis
Septic Shock
Pulmonary Infections
Empyema
Lung Abscess
Pneumonia
Disorders Of Airways
Acute COPD Exacerbation
COPD
Asthma
Obesity Hypoventilation Syndrome
Obstructive Sleep Apnea
Respiratory Failures
Acute Lung Injury
Acute Respiratory Distress
Acute Respiratory Failure
Metabolic Acidosis
Respiratory Acidosis
Patient
Demographics
Comorbidities
Cardiovascular Disorders
Endocrine & Metabolic Disorders
Neuromuscular Disorders
Cancer
Immunologic Disorders
Pulmonary Disorders
Renal Disorders
Outcome
Avoiding Intubation
Intubation Required
Reduction Hypercapnia
Weaning Failure
Weaning Successful
Therapy
Devices
High Flow Oxygen Systems
PAP Machines
Parameters
Oxygen Settings
Pressure Settings
Respiratory Rate
Volume Settings
Therapy Types
BiPAP
CPAP
HFNC
High Flow Mask
Timing
Duration
Interval Time
Time Event
Class
SubClass

**Fig 2** Class Hierarchy of the NIRS Ontology. BiPAP, Bilevel Positive Airway Pressure; CPAP, Continuous Positive Airway Pressure; COPD, Chronic Obstructive Pulmonary Disease; HFNC, High-Flow Nasal Cannula. Note: This figure displays only top-level classes. The hierarchical links represent semantic relationships and do not imply a deterministic clinical treatment pathway or causal relationship.

Similarly, the 'Outcome' superclass represents clinical endpoints of NIRS strategies, such as avoiding intubation or requiring escalation due to insufficient initial treatment. This hierarchical structure, with superclasses branching into detailed subclasses, provides a robust framework for modeling diverse respiratory care scenarios.

## Object and Data Properties and Conceptual Links

**Table 1** shows the key object properties defined within the NIRS ontology, indicating the relationships between domain and range classes. For example, the '*hasComorbidity*' property links patients with their comorbidities, '*hasDevice*' links therapies to their devices, and similarly, patients or therapies to specific outcomes are linked by the '*hasOutcome*' property. Each property is associated with specific domains and ranges. These object properties ensure that classes logically interact and accurately represent clinical relationships within respiratory support scenarios.

**Table 1**. NIRS Ontology Object Properties

| Object Property | Domain | Range |
| --- | --- | --- |
| hasComorbidity | Patient | Comorbidity |
| hasOutcome | Patient, TherapyType | Outcome |
| hasDevice | TherapyType | Device |
| hasTherapyType | Patient | TherapyType |
| hasRiskFactor | Patient | Comorbidity, Indication |
| hasParameter | TherapyType, Device | Parameter |
| isUsedIn | Device, Parameter | TherapyType |
| recommendedTherapyType | Patient | TherapyType |
| hasIndication | Patient | Indication |
| hasTiming | TherapyType | Timing |
| hasTreatmentEffect | Patient, TherapyType | Outcome |

NIRS ontology uses data properties to capture numeric or time-related details that are essential to clinical decision-making. These properties store attributes such as a patient's age using the data property name '*hasAge*', body mass index using '*hasBMI*', and ventilation settings using '*hasFiO2Value*', '*hasPEEPValue*', etc. Recording these details as data properties allows for logical reasoning in Protégé. **Table 2** shows the data property definitions, indicating their domain, such as for Patient or PEEP, and the associated data type, such as integer or decimal. This definition enables ontology to enforce logical restrictions on how data is stored and how it would be queried.

**Table 2**. NIRS Ontology Data Properties

| **Data Property** | **Domain** | **DataType** |
|---|---|---|
| hasAge | Age | Integer |
| hasBMI | BMI | Float |
| hasDuration | Duration | Float |
| hasSFRatio | Patient | Float |
| hasPFRatio | Patient | Float |
| hasEPAPValue | EPAP | Float |
| hasFiO2Value | $FiO_2$ | Float |
| hasIPAPValue | IPAP | Float |
| hasOxygenFlowRate | OxygenFlowRate | Float |
| hasPEEPValue | PEEP | Float |
| hasTIme | Timing | DateTime |
| hasRespiratoryRate | RespiratoryRate | Integer |

BMI, Body Mass Index; EPAP, Expiratory Positive Airway Pressure; $FiO_2$, Fraction of Inspired Oxygen; IPAP, Inspiratory Positive Airway Pressure; PEEP, Positive End-Expiratory Pressure.

**Fig 3** highlights the logical flow and relationships among ontology super classes. For instance, for a patient, the logical flow begins with the 'Patient' superclass, linking to clinical indications such as COPD through properties such as '*hasIndication*' or connecting to comorbidities such as Asthma or Bronchospasm via '*hasComorbidity*'. In addition, 'Patient' superclass captures individual attributes using data properties such as '*hasAge*' and '*hasBMI*'. The 'Therapy' superclass has therapeutic methods such as CPAP (continuous positive airway pressure) or devices such as BiPAP (bi-level positive airway pressure), identified by properties '*hasTherapyType*' and '*hasDevice*'.

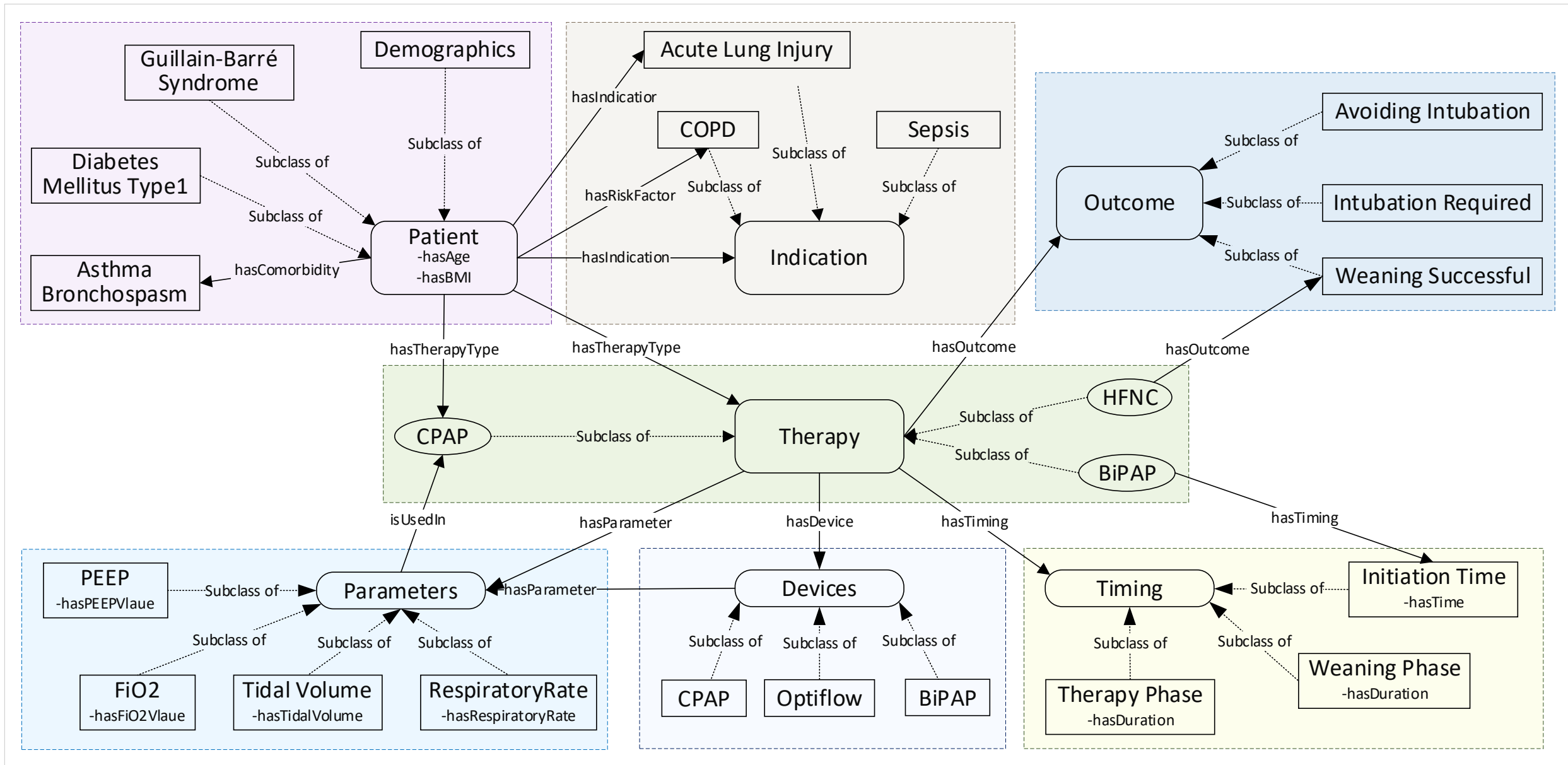

**Fig 3** Conceptual Model of the NIRS Ontology with Class Hierarchies and Property Relationships. BiPAP, Bilevel Positive Airway Pressure; CPAP, Continuous Positive Airway Pressure; COPD, Chronic Obstructive Pulmonary Disease; $FiO_2$, Fraction of Inspired Oxygen; HFNC, High-Flow Nasal Cannula; PEEP, Positive End-Expiratory Pressure. Note: The arrows indicate that a relationship can exist between concepts (e.g., a Patient record containing both an Indication and a Therapy), and do not imply a deterministic clinical treatment pathway or causal

relationship. Clinical decision logic and safety thresholds are handled separately by the SWRL rules described in Table 4.

The 'Therapy' class further specifies parameters such as PEEP using '*hasParameter*' and stores their values using data properties such as '*hasPEEPValue*' or '*hasFiO2Value*', and it captures timing details, such as 'Initiation Time' through '*hasTiming*'. Finally, the logical sequence culminates in the 'Outcome' class, which defines results of therapy such as "Weaning Successful" or "Intubation Required" via '*hasOutcome*'. This structured flow ensures that the ontology accurately reflects realistic clinical scenarios encountered in NIRS.

## Clinical Instances

We created clinical instances that are representative of practical scenarios from acute care settings by extracting examples of patient-ICU-stay from the eICU database to facilitate a precise evaluation of the NIRS ontology. Each scenario encompasses patient diagnoses, therapeutic interventions, clinical parameters, and outcomes, demonstrating the ontology's practical usability and analytic capability. A few example scenarios are illustrated in **Table 3,** where the Time Offset column captures the timing of clinical events in minutes relative to unit admission and the NIRS modalities or device assertions during the ICU stay.

**Table 3.** Example of Clinical Instances for Patients

| Patient | Time Offset | Property Assertion | Class Instance |
|---|---|---|---|
| Patient_01 | -11 | hasComorbidty | Type 2 Diabetes |
| | 12 | hasIndication | COPD Exacerbation |
| | 35 | hasDeviceType | BiPAP/CPAP |
| | 36 | hasPEEPValue | 5 cmH$_2$O |
| | 36 | hasFiO2Value | 60% |
| Patient_02 | 113 | hasIndication | Acute Respiratory Failure |
| | 1060 | hasFiO2Value | 55% |

| | 1060 | hasPaO2Value | 60 mm-Hg |
|---|---|---|---|
| | 1060 | haspHValue | 7.47 |
| | 1090 | hasFiO2Value | 55% |
| | 1190 | hasRespiratoryRate | 25 bpm |
| | 1250 | hasSpO2Value | 93% |
| | 1250 - 2870 | hasDeviceType | HFNC |
| Patient_03 | 19945 | hasRespiratoryRate | 23 bpm |
| | 20005 | hasSpO2Value | 96% |
| | 20033 | hasIndication | Pulmonary Edema |
| | 20033 | hasTherapyType | CPAP |
| Patient_04 | 1572 | hasIndication | ARDS |
| | 4212 | hasFiO2Value | 40% |
| | 4212 | hasPaO2Value | 80 mm-Hg |
| | 4270 | hasPEEPValue | 5 $cmH_2O$ |
| | 6090 | hasTherapyType | CPAP |
| Patient_05 | 85585 | hasIndication | Pneumonia |
| | 85585 | hasIndication | COPD Exacerbation |
| | 85920 | hasPEEPValue | 14 $cmH_2O$ |
| | 85950 | hasFiO2Value | 100% |
| | 86010 | hasFiO2Value | 100% |
| | 86030 | hasFiO2Value | 40% |
| | 86036 | hasPEEPValue | 15 $cmH_2O$ |
| | 86065 | hasOutcome | Discharged and expired or Dead/Weaning Failure |
| Patient_06 | 354 | hasPEEPValue | 12 $cmH_2O$ |
| | 445 | hasPaO2Value | 247 |
| | 445 | hasFiO2Value | 100% |
| | 598 | hasIndication | Sepsis |
| | 598 | hasTherapyType | CPAP |
| | 846 | hasFiO2Value | 60% |

ARDS, Acute Respiratory Distress Syndrome; COPD, Chronic Obstructive Pulmonary Disease; BiPAP, Bilevel Positive Airway Pressure; CPAP, Continuous Positive Airway Pressure; HFNC, High-Flow Nasal Cannula.

For example, Patient_01 has an indication of COPD Exacerbation with a comorbidity of Type 2 Diabetes and is treated with BiPAP/CPAP with a PEEP of 5 $cmH_2O$ and an $FiO_2$ of 60%. Separately, Patient_02 has an indication of Acute Respiratory Failure and is treated with HFNC with an $FiO_2$ of 55%, a respiratory rate of 25 bpm, and an $SpO_2$ of 93%. Similarly, Patient_04 has an indication of ARDS and is managed with CPAP, at a PEEP of 5 $cmH_2O$, with a P/F ratio of 200 mmHg ($FiO_2$ 40% and $PaO_2$ 80 mm-Hg, respectively).

## SWRL Rules and Clinical Inferences

SWRL rules were integrated into the ontology to enable automated clinical reasoning. They link clinical criteria to clinical instances for clearly defined competency questions and their answers (27,31). For example, according to the Berlin Definition of ARDS (acute respiratory distress syndrome), one SWRL rule in the NIRS ontology infers NIRS modalities (such as CPAP) as the recommended therapy for Mild ARDS patients with a P/F ratio between 200 and 300 mm-Hg and requiring a minimum PEEP of (≥5 $cmH_2O$) (32). While another rule, according to the Global Initiative for Chronic Obstructive Lung Disease( GOLD) 2024 guidelines, suggests that COPD patients with a $FiO_2$ > 40% should receive BiPAP as the first-line treatment for conditions like respiratory acidosis or cardiogenic pulmonary edema to reduce the risk of intubation and mortality (33). Similarly, according to the ATS Clinical Practice Guideline, patients with obesity hypoventilation syndrome (OHS) require CPAP as the first-line therapy (34). A selection of these competency questions and their corresponding rules is shown in **Table 4**. All other SWRL rules are described in the **S1 Appendix**.

A few rules are shown in **Table 4**. All other SWRL rules are described in the **S1 Appendix**.

**Table 4.** SWRL Rule for Competency Questionnaire and Inferred Outcome

| Competency Question | Clinical Instances | SWRL Rule | Inferred Outcome/Class |
|---|---|---|---|
| Which clinical indications lead to the initiation of NIRS, and what measurable criteria | If a Mild ARDS has a 200 ≤ P/F ratio ≤ 300 and PEEP ≥ 5, CPAP is recommended. | See rules in supporting information S1.1(Q1Rule 1) | Therapy Recommendation: CPAP |

| define when a patient meets those indications? | If a COPD patient requires $FiO_2 \geq 0.40$, BiPAP is recommended. | See rules in supporting information S1.1(Q1Rule 3) | Therapy Recommendation: BiPAP |
|---|---|---|---|
| How do the timing and duration of NIRS modalities influence patient outcomes? | If HFNC started within 3 hours, acute respiratory failure reduces intubation risk. | See rules in supporting information S1.1(Q3Rule 1) | Avoiding Intubation |
| | Weaning success likelihood increases when EPAP is reduced to 4 cmH2O, requiring a mean duration of 35 hours with BiPAP. | See rules in supporting information S1.1(Q3Rule 3) | Weaning Successful |

ARDS, Acute Respiratory Distress Syndrome; BiPAP, Bilevel Positive Airway Pressure; COPD, Chronic Obstructive Pulmonary Disease; $FiO_2$, Fraction of Inspired Oxygen; HFNC, High-Flow Nasal Cannula.

Defining these rules in SWRL allows the ontology to dynamically recommend therapies or identify patients at greater risk of adverse outcomes. For instance, initiating HFNC within three hours of detecting acute hypoxemic respiratory failure can help avert intubation (35).

On the other hand, prolonged BiPAP use beyond 24 to 48 hours in patients with elevated pressure support and PEEP (10 - 12 $cmH_2O$) may heighten the risk of intubation (33). This integration of rule-based logic demonstrates how ontology supports critical clinical judgments and fosters data-driven insights in respiratory care.

## Ontology Integration with Standardized Biomedical Terminologies

To enhance semantic alignment with established standards, the NIRS ontology incorporates structured mappings to standardized biomedical coding systems. Using BioPortal's search and cross-referencing tools, relevant classes were mapped to external vocabularies, such as ICD-10-CM, MedDRA, and SNOMED-CT. These mappings were performed by using annotations in Protégé for the leaf or terminal classes. A few examples of mappings, where each entry includes the class name, a descriptive label, an ontology code, an ontology link, and a brief comment highlighting its clinical use, are shown in **Table 5**. For instance, CPAP is mapped to HCPCS code E0601, referencing its role in maintaining constant airway pressure in non-invasive ventilation. BiPAP corresponds to a MedDRA code describing its dual-level pressure support. HFNC and High Flow Mask are linked to SNOMED-CT and MedDRA codes, respectively, both representing their function in delivering heated, humidified oxygen at high flow rates. These mappings strengthen interoperability, allowing ontology to integrate more effectively into clinical and research information systems.

**Table 5.** Mappings for NIRS therapy Type Classes

| **Class** | **Label** | **Ontology Code** | **Ontology Link** | **Comment** |
|---|---|---|---|---|
| CPAP | Continuous Positive Airway Pressure | HCPCS: E0601 | HCPCS | A type of non-invasive respiratory support (NIRS) used to maintain constant pressure in the airways throughout the respiratory cycle |
| BiPAP | Bilevel Positive Airway Pressure | MEDDRA: 10064530 | MedDRA | An NIRS ventilation mode delivering two pressure levels: a constant expiratory pressure (EPAP) and an increased pressure support for inspiration (IPAP). |
| HFNC | High Flow Nasal Cannula | SNOMED-CT: 426854004 | SNOMED-CT | A non-invasive respiratory support that delivers heated, humidified oxygen at high flow rates (up to 60 L/min) via nasal prongs. |
| HighFlowMask | High Flow Mask Oxygen Therapy | MEDDRA: 10084914 | MedDRA | A non-invasive respiratory support using heated, humidified oxygen at high flow rates (up to 15 L/min) to improve oxygenation. |

## Results:

The NIRS ontology consists of 145 classes (*Therapy*: 50, *Indication*: 25, *Patient*: 58, *Outcome*: 8), 11 object properties, 18 data properties, and 26 individuals (including 6 patient clinical scenarios). It was built upon 181 subclass axioms, supported by defined object property domains and ranges, as well as annotation properties such as '*rdfs:label'*, '*rdfs:comment'*, and external ontology codes. In total, 949 axioms were defined to capture semantic relationships among therapy parameters, clinical indications, comorbidities, timing, and outcomes.

To evaluate the reasoning ability to answer NIRS ontology's competency questions, we applied SWRL rules to a sample of 6 patients extracted from the eICU database. SPARQL queries retrieved inferred answers to those questions and extracted instances where patients met specific, measurable criteria (**Fig 4**). For example, a patient (Patient_01) with a COPD indication and an $FiO_2$ requirement of 60% inferred a recommendation for the NIRS modality of BiPAP. Similarly, a patient (Patient_04) with Mild ARDS, a P/F ratio of 200 mm-Hg, and a PEEP value of 5 $cmH_2O$, was inferred to require CPAP (see rules in supporting information S1.1)

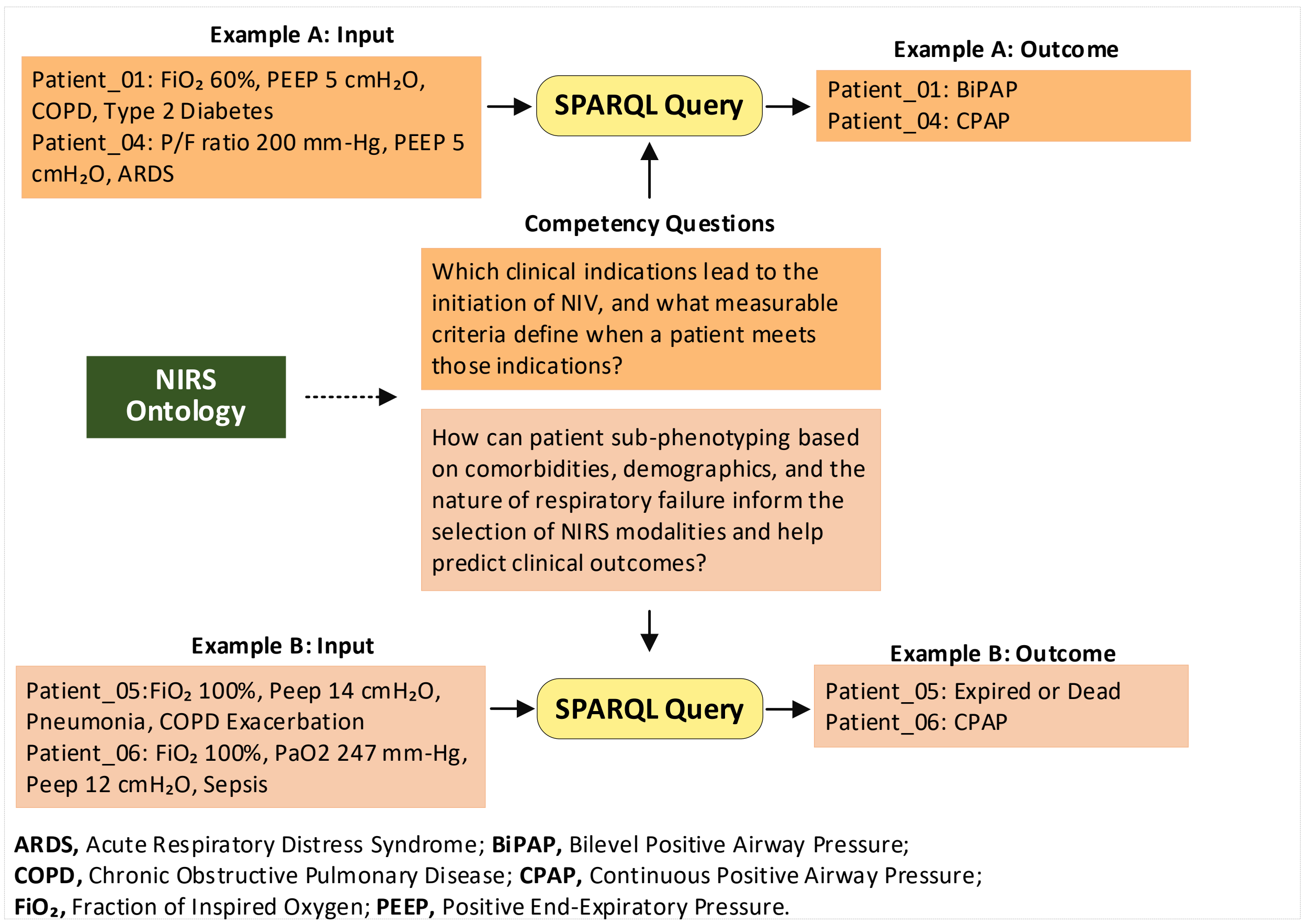

**Fig 4** Evaluation of SWRL rule logic for competency questionnaire. ('Example A or B: Input') Patient's scenario classes related to Competency question 1 or 2. (Example A or B: Outcome') Patient's scenario outcome classes retrieved by SPARQL.

Regarding risk level, a sepsis patient (Patient_06) with an $FiO_2$ value of 100% inferred a higher risk of intubation, while a patient (Patient_05) with COPD exacerbation requiring high $FiO_2$ of 100% with a P/F ratio of 247 mm-Hg (with a $PaO_2$ value of 247 mm-Hg) inferred an elevated mortality risk due to weaning failure (see rules in supporting information S1.2).

We also evaluated the ontology's capacity to support semantic interoperability. Out of 100 leaf or terminal classes, 87 were successfully mapped to standardized codes, a list of these mappings is provided in **S2 Table**. The remaining 13 classes, such as Oximizer, Optiflow, and Avoiding Intubation, represent domain-specific concepts that currently lack direct equivalents in standard terminologies. With the addition of annotations, the extracted outputs can include standardized vocabulary with codes and the descriptive definitions ('rdfs:comment') or full form ('rdfs:label') of different clinical concepts, depending on user needs. For instance, a patient (Patient_03) was recommended for BiPAP therapy with a COPD indication. In this case, the query may concern the details of COPD indications or BiPAP therapy, where COPD is annotated with the SNOMED-CT vocabulary using a concept code of 13645005, and BiPAP is annotated with MedDRA using a concept code of 10064530. Similarly, for a patient (Patient_06), who was at risk for intubation, the query can retrieve information about the risk of intubation or parameters with relevant vocabulary and concept codes (**Fig 5**)

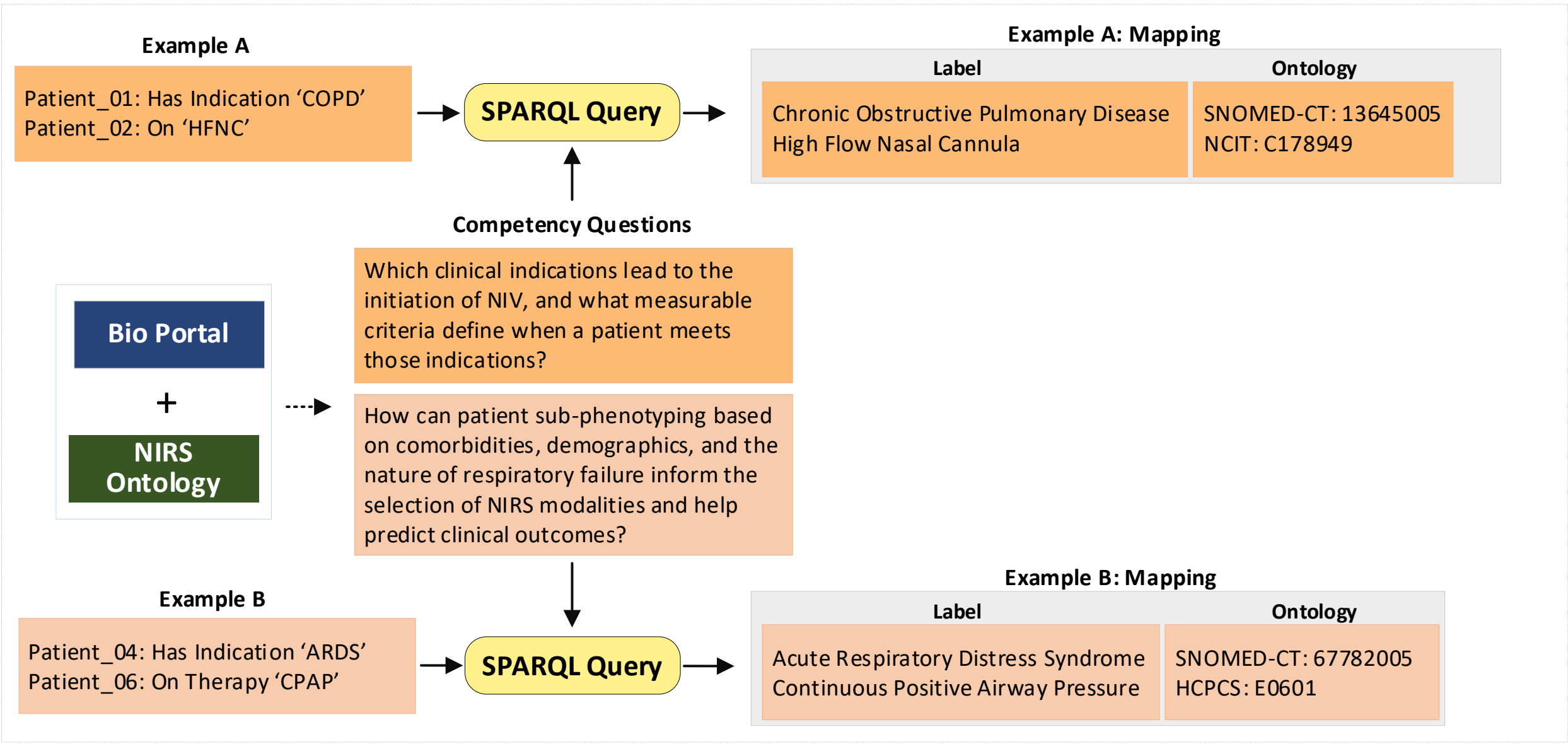

**Fig 5** Retrieving Mapping Annotations by SPARQL. (Example A) Patient's indications and outcomes classes related to Competency question 1. (Example B) Patient's indications and therapy type classes related to Competency question 2. (Example A or B: Mapping) Shows retrieved classes, labels, or definitions and Ontology mapping.

# Discussion

In this study, we developed the NIRS ontology to provide a unified framework for NIRS modalities and introduced a semantic reasoning layer. Existing ontologies, such as the Time Event Ontology (18), and other standardized ICU diagnoses (16), however, lack therapeutic parameters needed for phenotyping NIRS interventions. Our framework addresses this limitation by linking clinical concepts with rule-based inference. For example, whereas standard queries rely on explicit coding to identify "Respiratory Failure", the NIRS ontology uses SWRL rules to infer this physiological data (e.g., P/F ratio) and therapy settings.

The ontology design supports interoperability through both its structured class hierarchy and the use of annotation properties, such as 'rdfs:label', 'rdfs:comment', and 'hasOntoCode'. The classes were organized to reflect core clinical concepts such as indications, therapy types, outcomes, and device settings, making them compatible with standardized vocabularies. We mapped approximately 87% of leaf classes to external terminologies, while domain-specific concepts without direct equivalents (e.g., proprietary device names) were defined using internal annotations. Annotations were applied selectively based on the relevance of each class in each scenario, allowing targeted mapping without unnecessary duplication. For example, the outcome "Intubation Required" was mapped to a concept in the NCI Thesaurus, while the therapy type

BiPAP was linked to a MedDRA term. Mapping these terminal nodes to standard codes gives the NIRS ontology support for integration with diverse EHR settings.

While the existing classes and rules support therapy recommendations and outcomes, the scope of the NIRS ontology is limited to acute and decompensated chronic conditions. Furthermore, the current implementation features a set of 17 SWRL rules, a subset of which was validated against clinical instances to demonstrate technical feasibility. Additional rules are needed to support future questions involving more granular conditions or evolving therapy states. Expanding the ontology to include broader SWRL coverage would enhance its capacity to represent clinical complexity. However, a prospective evaluation of ontology is required to validate its effectiveness across different EHR systems.

## Limitation

Our study has several limitations. First, the definition of the SWRL rule relies on guideline-based clinical scenarios, with expert review by an intensivist. However, the ontology has not been tested with retrospective analysis to measure its predictive accuracy. Second, although our ontology captures basic temporal concepts (e.g., duration), it does not fully represent complex, continuous time-series trends such as rapid desaturation over minutes, which are vital for real-time monitoring. Third, the thresholds in our rules (e.g., $FiO_2 \geq 0.40$ for COPD) are derived from specific guidelines and may not align with local protocols or the clinical gray zones. Fourth, the ontology was internally validated using eICU data from 2014-2015. While the clinical concepts are still relevant, new modalities or interfaces developed after 2015 may require updates to the class hierarchy.

## Conclusion

This study provides a unified NIRS ontology that captures key components, including patient characteristics, clinical indications, therapy parameters, and outcomes. The ontology was developed using real-world eICU data and encoded with OWL semantics. It allows rule-based reasoning and patient-specific query execution through SWRL and SPARQL. The ontology facilitates a clear representation of therapy logic across various clinical contexts by incorporating NIRS modalities in a consistent, machine-readable structure. This NIRS framework demonstrates how complex, context-dependent therapies can be formally represented and queried, offering a foundation for more consistent documentation practices, integration into clinical data models, and advanced analysis of NIRS outcomes. However, while this work represents a foundational framework, it requires validation in future prospective studies.

## Author Contributions

CRediT author statement: Md Fantacher Islam: Data curation, formal analysis, investigation, methodology, software, visualization, writing – original draft. Jarrod Mosier: Supervision, validation, writing – review and editing. Vignesh Subbian: Conceptualization, project administration, methodology, supervision, resources, validation, writing – review and editing.

## Data Availability

The ontology files required to reproduce the results are available at the National Center for Biomedical Ontology (NCBO) Portal: https://bioportal.bioontology.org/ontologies/NIRS. The source repository is also hosted on GitHub: https://github.com/vsubbian/NIRS-Ontology

# Supporting Information

## S1 APPENDIX

### S1.1 SWRL Rules

This section describes the Semantic Web Rule Language (SWRL) rules that are designed to answer specific competency questions by enabling inference over the ontology.

#### S1.1.1 Competency Question (Q1):

*Which clinical indications lead to the initiation of noninvasive ventilation and what measurable criteria when a patient meets those indications?*

**Q1Rule 1:**

Rule Name: MildARDS_ModeratePFRatio_CPAP
Rule Guideline: 2012 Journal of the American Medical Association's Berlin Definition Report for ARDS (1)
Rule Comment: If a Mild ARDS has 200 ≤ P/F ratio ≤ 300 and PEEP/CPAP ≥ 5, CPAP is recommended.
Rule Code:

```
Patient(?pt) ^ hasIndication(?pt, AcuteRespiratoryDistress) ^ hasPFRatio(?pt, ?pfr) ^
swrlb:lessThanOrEqual(?pfr, 300) ^ swrlb:greaterThanOrEqual(?pfr, 200) ^
hasPEEPValue(?pt, ?peep) ^ swrlb:greaterThanOrEqual(?peep, 5) ->
recommendedTherapyType(?pt, CPAP)
```

**Q1Rule 2:**

Rule Name: NIRS_For_CHF_Respiratory_Distress
Rule Guideline: 2021 European Society of Cardiology Guidelines Report (2)
Rule Comment: For CHF patients with respiratory distress (RR >25 breaths/min, SpO2 <90%), NIRS modality is recommended to alleviate dyspnea and reduce the rate of mechanical endotracheal intubation.
Rule Code:

```
Patient(?pt) ^ hasIndication(?pt, CongestiveHeartFailure) ^ hasRespiratoryRate(?pt, ?rr) ^
swrlb:greaterThan(?rr, 25) ^ hasSpO2Value(?pt, ?spo2) ^ swrlb:lessThan(?spo2, 90) ->
recommendedTherapyType(?pt, NIRSModality)
```

**Q1Rule 3:**

Rule Name: COPD_HighFiO2_BiPAP
Rule Guideline: 2025 Global Initiative for Chronic Obstructive Lung Disease Report
Rule Comment: If a COPD patient requires $FiO_2$ ≥ 0.40, BiPAP is recommended.
Rule Code:

```
Patient(?pt) ^ hasIndication(?pt, COPD) ^ hasFiO2Value(?pt, ?fiO2) ^
swrlb:greaterThanOrEqual(?fiO2, 0.40) -> recommendedTherapyType(?pt, BiPAP)
```

**Q1Rule 4:**

Rule Name: HighIPAP_HighPEEP_Intubation

Rule Guideline: 2016 British Thoracic Society and Intensive Care Society Report (3)

Rule Comment: If high IPAP (>20–30 cm$H_2O$) or EPAP (>10–12 cm$H_2O$) are required without improvement in pH, $PaCO_2$, or respiratory rate despite optimization, the patient is at high risk of NIRS failure and needing intubation.

Rule Code:

```
Patient(?pt) ^ hasIPAPValue(?pt, ?ipap) ^ swrlb:greaterThan(?ipap, 20) ^
hasEPAPValue(?pt, ?epap) ^ swrlb:greaterThan(?epap, 10) -> hasOutcome(?pt,
IntubationRequired)
```

**Q1Rule 5:**

Rule Name: LowFiO2_LowPEEP_WeaningSuccess.

Rule Guideline: 2008 National Institutes of Health Network Mechanical Ventilation Protocol Summary

Rule Comment: Patients stable on $FiO_2 \leq 0.40$ and PEEP $\leq 8$ cm $H_2O$ are candidates for spontaneous breathing trials, indicating potential weaning success if sustained post-extubation.

Rule Code:

```
Patient(?pt) ^ hasFiO2Value(?pt, ?fiO2) ^ swrlb:lessThanOrEqual(?fiO2, 0.40) ^
hasPEEPValue(?pt, ?peep) ^ swrlb:lessThanOrEqual(?peep, 8) -> hasOutcome(?pt,
WeaningSuccessful)
```

**Q1Rule 6:**

Rule Name: NIRS_Failure_Leads_To_Intubation

Rule Guideline: 2021 Pulmonology Journal Non-invasive Respiratory Support Algorithm Report (4)

Rule Comment: If P/F Ratio < 100 or PaCO2 increases by 20% from basal levels, the patient is at risk of Non-invasive respiratory support failure and needs intubation.

Rule Code:

```
Patient(?pt) ^ hasPFRatio(?pt, ?pfr) ^ swrlb:lessThan(?pfr, 100) -> hasOutcome(?pt,
IntubationRequired)
```

**Q1Rule 7:**

Rule Name: ObesityHypoventilationSyndrome_BiPAP

Rule Guideline: 2012 New South Wales Agency and 2010 German Pneumology Society Guidelines (5)

Rule Comment: For OHS patients with hypercapnia (PaCO2 >45 mmHg), recommend BiPAP if CPAP fails to resolve hypoventilation.

Rule Code:

```
Patient(?pt) ^ hasIndication(?pt, ObesityHypoventilationSyndrome) ^ hasPaCO2(?pt,
?paco2) ^ swrlb:greaterThan(?paco2, 45) ^ hasTherapyType(?pt, CPAP) ->
recommendedTherapyType(?pt, BiPAP)
```

**Q1Rule 8:**

Rule Name: Pneumonia_Hypoxemia_HFNC (6)

Rule Guideline: 2024 Annals of Intensive Care Narrative Review
Rule Comment: For Pneumonia patients with hypoxemia (SpO2 <90% on >10 L/min standard O2) and high work of breathing (RR >25), recommend HFNC.
Rule Code:

```
Patient(?pt) ^ hasIndication(?pt, Pneumonia) ^ hasSpO2(?pt, ?spo2) ^
swrlb:lessThan(?spo2, 90) ^ hasOxygenFlowRate(?pt, ?flow) ^ swrlb:greaterThan(?flow,
10) ^ hasRespiratoryRate(?pt, ?rr) ^ swrlb:greaterThan(?rr, 25) ->
recommendedTherapyType(?pt, HFNC)
```

### S1.1.2 Competency Question (Q2):

How can patient sub-phenotyping based on comorbidities, demographics, and the nature of respiratory failure inform the selection of NIRS modalities and help predict clinical outcomes?

**Q2Rule 1:**
Rule Name: Elderly_COVID_NIRS_Failure
Rule Guideline: 2023 Internal and Emergency Medicine Report (7)
Rule Comment: Elderly (≥ 75 years) COVID-19 patients have a higher risk of NIRS failure or mortality.
Rule Code:

```
Patient(?pt) ^ hasIndication(?pt, COVID19) ^ hasAge(?pt, ?age) ^
swrlb:greaterThanOrEqual(?age, 75) -> hasOutcome(?pt, WeaningFailure)
```

**Q2Rule 2:**
Rule Name: Neuromuscular_NIRS_Modality
Rule Guideline: 2016 British Thoracic Society and Intensive Care Society Guideline Report (3)
Rule Comment: If a neuromuscular disease is diagnosed, the patient is recommended to the NIRS Modality.
Rule Code:

```
Patient(?pt) ^ hasIndication(?pt, NeuromuscularDisorders) ->
recommendedTherapyType(?pt, NIRSModality)
```

**Q2Rule 3:**
Rule Name: OHS_CPAP_FirstLine_Therapy
Rule Guideline: 2019 Official American Thoracic Society Clinical Practice Guideline (8)
Rule Comment: For Obesity Hypoventilation patients require CPAP as the first line therapy.
Rule Code:

```
Patient(?pt) ^ hasIndication(?pt, ObesityHypoventilationSyndrome) ->
recommendedTherapyType(?pt, CPAP)
```

**Q2Rule 4:**
Rule Name: Sepsis_NIV_Survival_Threshold
Rule Guideline: 2025 BioMed Central Pulmonary Medicine Report (9)
Rule Comment: Sepsis patients with PaO2/FiO2 ≥ 241 show increased survival rates and avoid intubation with initial NIRS modality.

Rule Code:

Patient(?pt) ^ hasIndication(?pt, Sepsis) ^ hasPFRatio(?pt, ?pfr) ^ swrlb:greaterThanOrEqual(?pfr, 241) ^ hasTherapyType(?pt, ?t) ^ TherapyType(?t) -> hasOutcome(?pt, AvoidingIntubation)

**Q2Rule 5:**

Rule Name: COPD_Oxygen_Target_Avoids_Acidosis

Rule Guideline: 2017 British Thoracic Society Emergency Oxygen Use Guideline (10)

Rule Comment: COPD exacerbations requiring high $FiO_2$ (>0.60) elevate mortality risk through worsened hypercapnia and acidosis.

Rule Code:

Patient(?pt) ^ hasIndication(?pt, COPD) ^ hasFiO2Value(?pt, ?fio2) ^ swrlb:greaterThan(?fio2, 0.60) -> hasOutcome(?pt, WeaningFailure)

### S1.1.3 Competency Question (Q3):

How do the timing and duration of NIRS modalities influence patient outcomes?

**Q3Rule 1:**

Rule Name Early_HFNC_Reduces_Intubation

Rule Guideline: 2015 New England Journal of Medicine High Flow Oxygen Report (11)

Rule Comment: If HFNC is started within 3 hours, it may reduce intubation risk in patients with P/F $\leq$ 200 mm Hg.

Rule Code:

Patient(?pt) ^ hasPFRatio(?pt, ?pfr) ^ swrlb:lessThanOrEqual(?pfr, 200) ^ hasTherapyType(?pt, HFNC) ^ hasTiming(?pt, ?time) ^ swrlb:lessThanOrEqual(?time, 3) -> hasOutcome(?pt, AvoidingIntubation)

**Q3Rule 2:**

Rule Name: Prolonged_BiPAP_Leads_To_Intubation

Rule Guideline: 2022 Scientific Reports Deep Learning Model (12)

Rule Comment: Intubation after BiPAP duration > 24 hours is associated with fewer Ventilator Free Days.

Rule Code:

Patient(?pt) ^ hasTherapyType(?pt, BiPAP) ^ hasDuration(?pt, ?duration) ^ swrlb:greaterThan(?duration, 24) -> hasOutcome(?pt, IntubationRequired)

**Q3Rule 3:**

Rule Name: BiPAP_Weaning_EPAP_Target

Rule Guideline: 2009 Indian Journal of Critical Care Medicine Report (13)

Rule Comment: Weaning success likelihood increases when EPAP is reduced to 4 cmH2O, typically requiring a mean duration of 35 hours with BiPAP.

Rule Code:

Patient(?pt) ^ hasTherapyType(?pt, BiPAP) ^ hasEPAPValue(?pt, ?epap) ^
swrlb:lessThanOrEqual(?epap, 4) ^ hasDuration(?pt, ?duration) ^
swrlb:greaterThanOrEqual(?duration, 35) -> hasOutcome(?pt, WeaningSuccessful)

**S1.1.4 Competency Question (Q4):**

How do key therapy parameters (e.g., IPAP, EPAP, $FiO_2$) change in response to a patient's evolving respiratory status?

**Q4Rule 1:**

Rule Name: OHS_EPAP_for_Hypoxemia
Rule Guideline: 2016 British Thoracic Society Guideline Report (3)
Rule Comment: For Obesity Hypoventilation Syndrome patients, titrating EPAP to 10-15 $cmH_2O$ recruits collapsed lungs and improves oxygenation.
Rule Code:

Patient(?pt) ^ hasIndication(?pt, ObesityHypoventilationSyndrome) ^ hasEPAPValue(?pt, ?epap)^ swrlb:lessThanOrEqual(?ipap, 10) ^ swrlb:greaterThan(?ipap, 15)
-> hasOutcome(?pt, ImprovementOxygenation)

**Q4Rule 2:**

Rule Name: Low_SF_Ratio_Predicts_Failure
Rule Guideline: 2022 Scientific Reports Pediatric Bi-level Positive Airway Pressure Failure Study (12)
Rule Comment: For patients with an S/F ratio < 264, the likelihood of intubation increases.
Rule Code:

Patient(?pt) ^ hasSFRatio(?pt, ?sfr) ^ swrlb:lessThan(?sfr, 264) -> hasOutcome(?pt, IntubationRequired)

**Q4Rule 3:**

Rule Name: Increase_IPAP_for_Hypercapnia
Rule Guideline: 2016 British Thoracic Society Guideline Report (14)
Rule Comment: Higher IPAP (typically 20-30 $cmH_2O$) improves $PaCO_2$ clearance and reduces the need for intubation in hypercapnic COPD patients.
Rule Code:

Patient(?pt) ^ hasIndication(?pt, COPD) ^ hasIPAPValue(?pt, ?ipap)^
swrlb:lessThanOrEqual(?ipap, 20) ^ swrlb:greaterThan(?ipap, 30) -> hasOutcome(?pt, AvoidingIntubation)

**S1.2 SPARQL Query**

This section presents SPARQL queries to retrieve and validate information from the ontology in response to the competency questions.

### S1.2.1 Query for Q1 Rule 1

```
PREFIX nirs: <http://www.semanticweb.org/fantacher/ontologies/2025/1/nirs#>

SELECT
  ?patient
  (STR(?pfr) AS ?PFRatio_Value)
  (STR(?peep) AS ?PEEP_Value)
  (?indication AS ?Indication_Type)
  (?therapy AS ?Recommended_Therapy)

WHERE {
  ?patient a nirs:Patient .
  ?patient nirs:hasIndication nirs:AcuteRespiratoryDistress .

  # Match the rule's specific data points
  ?patient nirs:hasPFRatio ?pfr .
  ?patient nirs:hasPEEPValue ?peep .

  OPTIONAL { ?patient nirs:hasIndication ?indication . }
  OPTIONAL { ?patient nirs:recommendedTherapyType ?therapy . }

  # Filter based on Q1Rule1 logic: 200 <= PFR <= 300 AND PEEP >= 5
  FILTER (?pfr >= 200 && ?pfr <= 300 && ?peep >= 5)
}
```

### S1.2.2 Query for Q1 Rule 3

```
PREFIX nirs: <http://www.semanticweb.org/fantacher/ontologies/2025/1/nirs#>

SELECT ?patient
      (?indication AS ?Indication_Type)
      (STR(?fio2) AS ?FiO2_Value)
      (?therapy AS ?Recommended_Therapy)

WHERE {
  ?patient a nirs:Patient .
  ?patient nirs:hasIndication nirs:COPD .

  OPTIONAL { ?patient nirs:hasIndication ?indication . }
  OPTIONAL { ?patient nirs:hasFiO2Value ?fio2. }
  OPTIONAL { ?patient nirs:recommendedTherapyType ?therapy . }

  FILTER(?fio2 >= 0.40)
}
```

**S1.2.3 Query for Q2 Rule 4**

```
PREFIX nirs: <http://www.semanticweb.org/fantacher/ontologies/2025/1/nirs#>

SELECT
  ?patient
  (STR(?pfr) AS ?PFRatio_Value)
  (?currentTherapy AS ?Current_Therapy)
  (?indication AS ?Indication_Type)
  (?outcome AS ?Predicted_Outcome)

WHERE {
  ?patient a nirs:Patient .
  ?patient nirs:hasIndication nirs:Sepsis .

  ?patient nirs:hasPFRatio ?pfr .
  ?patient nirs:hasTherapyType ?currentTherapy .

  # UPDATED LOGIC MATCHING SWRL: TherapyType(?t)
  # This checks that the therapy is an instance of the TherapyType class
  # (Reasoning must be enabled for this to catch subclasses like CPAP/BiPAP)
  ?currentTherapy a nirs:TherapyType .

  OPTIONAL { ?patient nirs:hasIndication ?indication . }
  OPTIONAL { ?patient nirs:hasOutcome ?outcome . }

  # Filter based on Q2Rule4 logic: Only PFR >= 241 is needed here
  FILTER (?pfr >= 241)
}
```

**S1.2.4 Query for Q2 Rule 5**

```
PREFIX nirs: <http://www.semanticweb.org/fantacher/ontologies/2025/1/nirs#>

SELECT
  ?patient
  (STR(?fio2) AS ?FiO2_Value)
  (?indication AS ?Indication_Type)
  (?outcome AS ?Predicted_Outcome)

WHERE {
  ?patient a nirs:Patient .
  ?patient nirs:hasIndication nirs:COPD .

  ?patient nirs:hasFiO2Value ?fio2 .

  OPTIONAL { ?patient nirs:hasIndication ?indication . }
```

```
  OPTIONAL { ?patient nirs:hasOutcome ?outcome . }

  # Filter based on Q2Rule5 logic: FiO2 > 0.60
  FILTER (?fio2 > 0.60)
}
```

**S1.3 SPARQL Query with RDFS Mapping Annotations**

This section presents SPARQL queries designed to retrieve and verify annotations assigned to ontology classes.

**S1.2.1 Query with Annotations for Q1 Rule 1**

```
PREFIX rdfs: <http://www.w3.org/2000/01/rdf-schema#>
PREFIX nirs: <http://www.semanticweb.org/fantacher/ontologies/2025/1/nirs#>

SELECT
  ?patient
  (STR(?pfr) AS ?PFRatio_Value)
  (STR(?peep) AS ?PEEP_Value)
  (?therapy AS ?Recommended_Therapy)
  (?indication AS ?Indication_Type)
  (STR(?label) AS ?Label)
   (STR(?ontologyCode) AS ?OntologyCode)
  (STR(?comment) AS ?Comment)

WHERE {
  ?patient a nirs:Patient .
  ?patient nirs:hasIndication nirs:AcuteRespiratoryDistress .

  # Match the rule's specific data points
  ?patient nirs:hasPFRatio ?pfr .
  ?patient nirs:hasPEEPValue ?peep .

  OPTIONAL { ?patient nirs:hasIndication ?indication . }
  OPTIONAL { ?patient nirs:recommendedTherapyType ?therapy . }

  OPTIONAL { ?indication rdfs:label ?label . }
  OPTIONAL { ?indication nirs:hasOntoCode?ontologyCode . }
  OPTIONAL { ?indication rdfs:comment ?comment . }

  # Filter based on Q1Rule1 logic: 200 <= PFR <= 300 AND PEEP >= 5
  FILTER (?pfr >= 200 && ?pfr <= 300 && ?peep >= 5)
}
```

This section presents SPARQL queries designed to retrieve and verify annotations assigned to ontology classes.

**S1.2.1 Query with Annotations for Q1 Rule 3**

```
PREFIX rdfs: <http://www.w3.org/2000/01/rdf-schema#>
PREFIX nirs: <http://www.semanticweb.org/fantacher/ontologies/2025/1/nirs#>

SELECT ?patient
       (STR(?fio2) AS ?FiO2_Value)
       (?therapy AS ?Recommended_Therapy)
       (?indication AS ?Indication_Type)
       (STR(?label) AS ?Label)
       (STR(?ontologyCode) AS ?OntologyCode)
       (STR(?comment) AS ?Comment)

WHERE {
  ?patient a nirs:Patient .
  ?patient nirs:hasIndication nirs:COPD .

  OPTIONAL { ?patient nirs:hasIndication ?indication . }
   OPTIONAL { ?patient nirs:hasFiO2Value ?fio2. }
  OPTIONAL { ?patient nirs:recommendedTherapyType ?therapy . }

   OPTIONAL { ?indication rdfs:label ?label . }
   OPTIONAL { ?indication nirs:hasOntoCode?ontologyCode . }
   OPTIONAL { ?indication rdfs:comment ?comment . }

  FILTER(?fio2 >= 0.40)
}
```

**S1.3.3 Query with Annotations for Q2 Rule 4**

```
PREFIX rdfs: <http://www.w3.org/2000/01/rdf-schema#>
PREFIX nirs: <http://www.semanticweb.org/fantacher/ontologies/2025/1/nirs#>

SELECT
  ?patient
  (STR(?pfr) AS ?PFRatio_Value)
  (?currentTherapy AS ?Current_Therapy)
  (?outcome AS ?Predicted_Outcome)
  (?indication AS ?Indication_Type)
  (STR(?label) AS ?Label)
  (STR(?ontologyCode) AS ?OntologyCode)
  (STR(?comment) AS ?Comment)
```

```
WHERE {
  ?patient a nirs:Patient .
  ?patient nirs:hasIndication nirs:Sepsis .

  ?patient nirs:hasPFRatio ?pfr .
  ?patient nirs:hasTherapyType ?currentTherapy .

  # UPDATED LOGIC MATCHING SWRL: TherapyType(?t)
  # This checks that the therapy is an instance of the TherapyType class
  # (Reasoning must be enabled for this to catch subclasses like CPAP/BiPAP)
  ?currentTherapy a nirs:TherapyType .

  OPTIONAL { ?patient nirs:hasIndication ?indication . }
  OPTIONAL { ?patient nirs:hasOutcome ?outcome . }

  OPTIONAL { ?indication rdfs:label ?label . }
  OPTIONAL { ?indication nirs:hasOntoCode?ontologyCode . }
  OPTIONAL { ?indication rdfs:comment ?comment . }

  # Filter based on Q2Rule4 logic: Only PFR >= 241 is needed here
  FILTER (?pfr >= 241)
}
```

### S1.3.4 Query with Annotations for Q2 Rule 5

```
PREFIX rdfs: <http://www.w3.org/2000/01/rdf-schema#>
PREFIX nirs: <http://www.semanticweb.org/fantacher/ontologies/2025/1/nirs#>

SELECT
  ?patient
  (STR(?fio2) AS ?FiO2_Value)
  (?outcome AS ?Predicted_Outcome)
  (?indication AS ?Indication_Type)
  (STR(?label) AS ?Label)
  (STR(?ontologyCode) AS ?OntologyCode)
  (STR(?comment) AS ?Comment)

WHERE {
  ?patient a nirs:Patient .
  ?patient nirs:hasIndication nirs:COPD .

  ?patient nirs:hasFiO2Value ?fio2 .

  OPTIONAL { ?patient nirs:hasIndication ?indication . }
```

```
  OPTIONAL { ?patient nirs:hasOutcome ?outcome . }

  OPTIONAL { ?indication rdfs:label ?label . }
  OPTIONAL { ?indication nirs:hasOntoCode?ontologyCode . }
  OPTIONAL { ?indication rdfs:comment ?comment . }

  # Filter based on Q2Rule5 logic: FiO2 > 0.60
  FILTER (?fio2 > 0.60)
}
```

**S2 Table. Mapping of NIRS Ontology Leaf or Terminal Nodes**

| Superclass | Leaf/End Class (Concept) | Standard Vocabulary | Concept Code |
|---|---|---|---|
| Indication | Pulmonary Edema | SNOMEDCT | 19242006 |
| | AECOPD | SNOMEDCT | 195951007 |
| | Asthma | SNOMEDCT | 195967001 |
| | COPD | SNOMEDCT | 13645005 |
| | Obesity Hypoventilation Syndrome (OHS) | ICD9CM | 278.03 |
| | Obstructive Sleep Apnea (OSA) | SNOMEDCT | 78275009 |
| | COVID-19 | SNOMEDCT | 840539006 |
| | Empyema | SNOMEDCT | 312682007 |
| | Lung Abscess | SNOMEDCT | 73452002 |
| | Pneumonia | SNOMEDC | 233604007 |
| | Acute Lung Injury | NCIT | C155766 |
| | Acute Respiratory Distress Syndrome (ARDS) | SNOMEDCT | 67782005 |
| | Hypercapnic Respiratory Failure | SNOMEDCT | 709109004 |
| | Hypoxemic Respiratory Failure | SNOMEDCT | 10676831000119101 |
| | Mixed Respiratory Failure | N/A | N/A |
| | Metabolic Acidosis | SNOMEDCT | 59455009 |
| | Respiratory Acidosis | SNOMEDCT | 12326000 |
| | Sepsis | SNOMEDCT | 91302008 |
| | Septic Shock | SNOMEDCT | 76571007 |
| Outcome | Avoiding Intubation | N/A | N/A |
| | Intubation Required | NCIT Code | C178949 |
| | Weaning Failure | MEDDRA | 10066829 |
| | Successful Weaning | SNOMEDCT | 404998008 |
| | Improvement Oxygenation | SNOMEDCT | 284032008 |
| | Reduction Hypercapnia | SNOMEDCT | 29596007 |
| Patient | PaO2/FiO2 Ratio | IOBC | 201406062517222943 |
| | SpO2/FiO2 Ratio | N/A | N/A |
| | Hematologic Malignancy | DOID | 2531 |
| | Lung Cancer | LNC | LA15687-9 |
| | Metastatic Lung Cancer | LNC | LA28289-9 |
| | Thoracic Cancer | DOID | 5093 |
| | Body Mass Index (BMI) | LNC | 39156-5 |
| | Smoking Status | SNOMEDCT | 308512009 |
| | Age | N/A | N/A |
| | Sex | N/A | N/A |
| | Hematologic Malignancy | DOID | 2531 |
| | Lung Cancer | LNC | LA15687-9 |
| | Metastatic Lung Cancer | LNC | LA28289-9 |
| | Thoracic Cancer | DOID | 5093 |
| | Biventricular | ICD10CM | I50.82 |
| | CHF Diastolic | ICD10CM | I50.3 |
| | CHF Left Sided | ICD10CM | I50.1 |
| | CHF Right Sided | ICD10CM | I50.81 |
| | CHF Systolic | ICD10CM | I50.2 |
| | Cor-Pulmonale | SNOMEDCT | 83291003 |
| | Hypertension Heart Failure | SNOMEDCT | 84114007 |
| | Valvular Regurgitation | SNOMEDCT | 10337008 |
| | Valvular Stenosis | NCIT | C62433 |
| | DiabetesMellitusType1 | ICD10CM | E10 |
| | DiabetesMellitusType2 | ICD10CM | E11 |
| | AIDS | SNOMEDCT | 62479008 |
| | Immunosuppression | SNOMEDCT | 38013005 |
| | Sarcoidosis | SNOMEDCT | 31541009 |
| | SLE | SNOMEDCT | 55464009 |
| | Vasculitis | SNOMEDCT | 31996006 |
| | ALS | SNOMEDCT | 86044005 |
| | Guillain-Barre Syndrome | ICD10 | G61.0 |
| | Multiple Sclerosis | SNOMEDCT | 24700007 |
| | Neuropathy | SNOMEDCT | 386033004 |
| | Post-Polio Syndrome | MEDDRA | 10036239 |
| | Chronic Alcoholic Myopathy | SNOMEDCT | 838376007 |
| | Malignant Hyperthermia | SNOMEDCT | 405501007 |
| | Muscular Dystrophy | SNOMEDCT | 73297009 |
| | Myopathy Critical-Illness | SNOMEDCT | 443819006 |
| | Rhabdomyolysis | SNOMEDCT | 240131006 |
| | Asthma Bronchospasm | SNOMEDCT | 195967001 |
| | Asthma Mild | SNOMEDCT | 370218001 |
| | Pleural Effusion CHF | SNOMEDCT | 90727007 |
| | Diabetes Insipidus | ICD10CM | E23.2 |
| | Acute Kidney Injury | SNOMEDCT | 14669001 |
| | Chronic Kidney Disease | SNOMEDCT | 709044004 |
| | HFNC | SNOMEDCT | 426854004 |
| | Optiflow | N/A | N/A |
| | Oximizer | N/A | N/A |
| | Vapotherm | N/A | N/A |
| | BiPAP | MEDDRA | 10064530 |
| | CPAP | HCPCS | E0601 |
| | PaCO2 | SNOMEDCT | 25284008 |
| | pH | CMO | 0000379 |
| | TotalCO2 | OCHV | C0860709 |
| | FiO2 | LNC | LP286756-4 |
| | Oxygen Flow Rate | SNOMEDCT | 42708100 |
| | EPAP | SNOMEDCT | 699113009 |
| | IPAP | RCTV2 | 8726.00 |
| Therapy | PEEP | SNOMEDCT | 250854009 |
| | Pressure Support | LNC | LP73175-9 |
| | Set Respiratory Rate | SNOMEDCT | 86290005 |
| | Spontaneous Respiratory Rate | SNOMEDCT | 271625008 |
| | Inspiratory Flow Rate | LNC | LP101893-8 |
| | Inspiratory Pressure Set | SNOMEDCT | 417071008 |
| | Tidal Volume | LNC | LP188693-8 |
| | Acute Resuscitation | SNOMEDCT | 439569004 |
| | Ceiling Of Therapy | N/A | N/A |
| | Do Not Intubate | N/A | N/A |
| | Post-Extubation Support | SNOMEDCT | 718089001 |
| | Weaning Support | N/A | N/A |

| | | |
|---|---|---|
| High Flow Mask | MEDDRA | 10084914 |
| Cumulative Time | N/A | N/A |
| Therapy Duration | SNOMEDCT | 261773006 |
| Therapy Interval | SNOMEDCT | 261774000 |
| Weaning Interval | N/A | N/A |
| Initiation Time | LNC | MTHU046569 |
| Pause Time | SNOMEDCT | 250821007 |
| Termination Time | SNOMEDCT | 397898000 |

**SWRL Rule References**